\documentclass[letterpaper]{article} 
\usepackage{aaai25}  
\usepackage{times}  
\usepackage{helvet}  
\usepackage{courier}  
\usepackage[hyphens]{url}  
\usepackage{graphicx} 
\usepackage{natbib}  
\usepackage{caption} 
\usepackage{algorithm}
\usepackage{algorithmic}
\usepackage{amsmath} 
\usepackage{bm} 
\usepackage{amsfonts} 
\usepackage{multirow}
\usepackage{subcaption}
\usepackage{newfloat}
\usepackage{listings}

\DeclareCaptionStyle{ruled}{labelfont=normalfont,labelsep=colon,strut=off} 
\floatstyle{ruled}
\newfloat{listing}{tb}{lst}{}
\floatname{listing}{Listing}
\title{Algorithmic Contract Design with Reinforcement Learning Agents}
\author {
    David Molina Concha\textsuperscript{\rm 1},
   Kyeonghyeon Park\textsuperscript{\rm 2},
    Hyun-Rok Lee\textsuperscript{\rm 3},
     Taesik Lee\textsuperscript{\rm 2},
      Chi-Guhn Lee\textsuperscript{\rm 1}
}
\affiliations {
    \textsuperscript{\rm 1}University of Toronto\\
    \textsuperscript{\rm 2}Korea Advanced Institute of Science and Technology\\
    \textsuperscript{\rm 3}Inha University\\
    david.molina@mail.utoronto.ca, kyeonghyeon.park@kaist.ac.kr, hyunrok.lee@inha.ac.kr, taesik.lee@kaist.ac.kr, cglee@mie.utoronto.ca
}

\usepackage{bibentry}

\begin{document}

\maketitle

\begin{abstract}
We introduce a novel problem setting for algorithmic contract design, named the principal-MARL contract design problem. This setting extends traditional contract design to account for dynamic and stochastic environments using Markov Games and Multi-Agent Reinforcement Learning. To tackle this problem, we propose a Multi-Objective Bayesian Optimization (MOBO) framework named Constrained Pareto Maximum Entropy Search (cPMES). Our approach integrates MOBO and MARL to explore the highly constrained contract design space, identifying promising incentive and recruitment decisions. cPMES transforms the principal-MARL contract design problem into an unconstrained multi-objective problem,  leveraging the probability of feasibility as part of the objectives and ensuring promising designs predicted on the feasibility border are included in the Pareto front. By focusing the entropy prediction on designs within the Pareto set, cPMES mitigates the risk of the search strategy being overwhelmed by entropy from constraints. We demonstrate the effectiveness of cPMES through extensive benchmark studies in synthetic and simulated environments, showing its ability to find feasible contract designs that maximize the principal's objectives. Additionally, we provide theoretical support with a sub-linear regret bound concerning the number of iterations.
\end{abstract}

%

\section{Introduction}

In multi-agent contract design problems, a principal aims to align the incentives of individual agents with system-level objectives by designing contracts. Applications can range from data offloading using UAVs \cite{chen2022deepuav} to government contracts for social welfare \cite{dai2021contractforest}. Most of the literature assumes that system outcomes strictly depend on agents' actions, without considering their interaction with the environment. This assumption is difficult to maintain in real-life scenarios, where individual and system-level outcomes are often subject to external factors that cannot be perfectly controlled or predicted by the principal or agents \cite{weberxiong2006externalitiescd}.

External stochasticity can lead contracts to not yield the expected outcomes from the multi-agent system, prompting recent research  to account for dynamic scenarios \cite{xie2023behaviormulti,chen2022deepuav,zhao2023deep}. Markov Games (MGs) are a natural approach to modeling multi-agent systems in stochastic environments, capturing the interactions among multiple agents in dynamic environments. However, each contract design evaluation requires solving the MG using Multi-Agent Reinforcement Learning (MARL), which faces significant computational challenges due to the exponential increase in state and action space relative to the number of agents \cite{qu2020scalable}, making it infeasible to evaluate every design. Additionally, accounting for the contract feasibility constraints for each agent leads to a highly constrained problem \cite{chen2022deepuav}. Therefore, extending principal-multi-agent contract design problems to MGs remains an open challenge.

Bayesian Optimization (BO) has been applied in algorithmic reward design within MGs \cite{mguni2019coordinating, shou2020reward}, treating the outcomes of learned policies from MARL as expensive-to-evaluate black-box functions. However, extending these methods for contract design requires additional efforts to handle the set of feasibility constraints for the individual agents.  We propose a Multi-Objective BO (MOBO) framework to optimize contract designs in the presence of multiple Reinforcement Learning (RL) agents. Our approach, constrained Pareto Maximum Entropy Search (cPMES), is composed of two levels: the design optimization level and the MARL level. At the first level, we introduce the feasibility of the design as one of the objectives to be optimized \cite{yangetal2021constraintevolu}, obtaining a Pareto front of non-dominated solutions that accounts for the trade-off between the principal's objective and the probability of feasibility. Then, we prioritize designs based on the predictive information gain of the objective function and constraints. At the MARL level, cPMES models the selected design as a partially observable Multi-Type Markov Game (POMTMG) and solves it using MARL. The contributions of this work are:
\begin{itemize}
\item We extend the principal-multi-agent contract design problem to MGs and introduce the principal-MARL contract design problem.
\item We introduce cPMES, an efficient framework for solving the principal-MARL contract design problem.
\item We establish a sub-linear regret bound with respect to the number of iterations, supporting the effectiveness of our approach.
\item We showcase the performance of the cPMES framework through benchmark studies in synthetic and simulated environments with relevant methods in constrained BO.
\end{itemize}

\section{Related work}


Recent work in contract theory has explored various settings for the principal-multi-agent contract design problem \cite{dutting2023multilincontract,castiglioni2023indoutcomesmulti,xie2023behaviormulti}, however they assume that the system's outcomes depend solely on the agents' actions, and the recruitment of agents is not considered a decision variable.  Integration of algorithmic contract design and RL has been primarily explored in the context of single-agent systems. The work by \cite{chen2022deepuav} is only applicable to stochastic energy harvesting with UAVs and does not consider MGs. Recently, \cite{zhao2023deep} addressed principal-multi-agent contract design by modeling the problem as a Markov Decision Process (MDP) from principal's perspective, treating the multiple agents decisions as part of the environment.

Related problems, such as the incentive design problem, have been addressed in the domain of MGs and MARL. In the work of \cite{mguni2019coordinating, shou2020reward}, a bi-level optimization framework is proposed, using BO to optimize a single reward parameter and MARL to solve the underlying MG. However, their work does not consider incentive feasibility constraints, and recruitment is not part of the principal's decision variables. Introducing constrains to expensive to evaluate black-box-functions has led the research community to develop constrained BO methods that can be  categorized as probability of feasibility based methods, step look ahead methods and surrogate-assisted constraint handling methods \cite{wang2023recentbo}. Relevant methods in this field such as Expected Improvement (cEI), constrained Max-value Entropy Search   (cMES)~\cite{perrone2019constrained} and MACE~\cite{zhang2022efficient} struggle to optimize in highly constrained problems, suffering from lack of sampling close to the feasibility border and search strategies driven by the feasibility rather than the objective function ~\cite{bagheri2017constrainteicrit}.

\section{Preliminaries}

\subsection{Principal-Multi-Agent Contracts}

We consider the definition of principal-multi-agent problem provided in~\cite{castiglioni2023indoutcomesmulti}, where the contract design is defined by the tuple $<N, \Omega, A>$, where $N$ is the finite set of agents in the system, $\Omega$ is a finite set of all possible individual outcomes of each agent and $A$ a finite set of agents' actions. The principal aims to maximize their expected utility by committing to a contract, which specifies payments from the principal to each agent contingently
on the actual individual outcome achieved by the agent. The incentives of each agent can be expressed as a linear combination  by introducing an individual incentive weight $\alpha_i$~\cite{dutting2023multilincontract}. The weight symbolizes the proportion of return of principal that is awarded to each agent, then the utility of agent $i$ becomes:
\begin{align}
    r_i=\alpha_i * G_{\mathbf{a}} - \mathbf{l}c_i,
\end{align}

\noindent where $G_{\mathbf{a}}$ is the principal's objective, $\mathbf{a}$ is a tuple containing agents' actions,  $\mathbf{l}$  is a binary indicator function that outputs $1$ if agent $i$ exerted effort, incurring in a cost $c_i$, and $0$ in any other case. The incentive weight is not the only design variable that has been considered in this type of problems. One of the main motivations of \cite{BABAIOFF2012combin} is to analyse the impacts on the recruitment policy in terms of the contribution to the task given some contract. In this work, we consider the recruitment of agents as part of the decision variable of the contract design problem.

\subsection{Partially Observable Multi-Type  Markov Game}

In the presence of heterogeneous agents, the incorporation of multiple types to capture agent diversity has proven successful in the realm of MARL~\cite{subramanian2020multitype}. By extending the formulation of partially observable MG (POMG) to multiple types of agents, we can model the problem as a partially observable Multi-Type  Markov Game (POMTMG), defined by the tuple:
\begin{align}
<M, N, S, O, \{\sigma_i\}, A, \{r_i\}, T, \gamma>,
\end{align}

where $M$ is the number of available types for the agents and $N$ is the set of agents. Each team $\mathbf{n} \in N$ is an $M$-dimensional vector, where each entry in the vector is the number of agents in the type. $S$ is the joint state space of the $N$ agents and $O \equiv O_1 \times \ldots \times O_N$ is the joint observation space,  with $O_i \subseteq S$ for all $i \in N$. The joint action space is defined as  $A \equiv A_1 \times \ldots \times A_N$, $\{\sigma_i\}$ and $\{r_i\}$ represent the observation and reward functions for all $i \in N$, where $\sigma_i : S \rightarrow O_i$ and $r_i : S \times A \times S \rightarrow \mathbb{R}$, respectively. The model follows an stochastic state transition $T$, given by $T: S \times A \times S \rightarrow [0, 1]$. The discount factor is represented as $\gamma$. 

This concept enables us to effectively use MARL to train heterogeneous teams of agents with different incentives and/or actions spaces.

\section{Problem definition}

The principal aims to modify a baseline situation to maximize system level performance by defining a set of linear of contracts that considers incentives or penalties for a set of $N$ agents. For this problem, we consider the number of agents as part of the principal's contract design decisions \cite{BABAIOFF2012combin}. The modification of the number of agents is done by adding $N_a$ new agents to the system, assuming baseline agents,$N_b$, can not be fired. Therefore, $N_b \subseteq N, N_a \subseteq N, N_a \cap N_b = \emptyset, N_a \cup N_b = N$.  Then, the decision variables of the principal are the linear weight for incentive or penalty of each agent $\bm{\alpha}=[\alpha_1,..,\alpha_N]$ and the additional agents $N_a$. 

Inspired by the contract design problem setting  of~\cite{castiglioni2023indoutcomesmulti}, we assume principal can observe the individual outcomes of each agent and, given a contract, they behave according to:
\begin{enumerate}
    \item Best Response Strategies (BRS): the best response strategy for an agent in the context of MARL is a policy that maximizes its expected cumulative reward given the policies of the other agents:
    \begin{align}
        \label{eq:BR}
        \mathbb{E}_{\pi_i,\pi_{-i}}[r_i] \geq \mathbb{E}_{{\pi'}_i,\pi_{-i}}[r_i]
        \\
        \forall i \in \{1,\cdots,N\}, \forall \pi_i, \pi'_{i} \in \Pi_i, \pi_{-i} \in \Pi_{-i}, \nonumber
    \end{align}
    
    \noindent where $\pi_{-i}$ symbolizes the policy of all other agents except $i$. The policy $\pi_i$ is considered best response strategy if provides a greater or equal expected return than any other policy available for agent $\pi'_{i}$.

    \item Individual Rationality (IR): this principle ensures that agents have an incentive to participate in the contract. It implies that each agent's utility under the contract should be at least as much as their baseline situation:
    \begin{align}
        \label{eq:IR1}
        \mathbb{E}_{\pi_j^{\bm{\alpha},N_a},\pi_{-j}^{\bm{\alpha},N_a}}[r_j^{\alpha_j}] \geq  \mathbb{E}_{\pi_j,\pi_{-j}}[r_j]  
        \\  
        \forall j \in \{1,\cdots,N_b\}, \nonumber
    \end{align}
    
    \noindent the incentive parameter for agent $j$ is $\alpha_j$ and $\bm{\pi} = [\pi_1, ..., \pi_N]$. Note that the policy of the agents $\pi_j^{\bm{\alpha},N_a}$ and the reward functions are affected by the contract design decisions from principal. 

    Since the additional agents lack baseline utility for comparison, we set a minimum expected return $c$ to assess the IR principle as follows:
    \begin{align}
        \label{eq:IR2}  
        \mathbb{E}_{\bm{\pi}^{\bm{\alpha},N_a}}[r_k^{\alpha_k}] \geq  c 
        \\  
        \forall k \in \{1,\cdots,N_a\}, \nonumber
    \end{align}    
\end{enumerate}

From principal's perspective, the linear contract is feasible only if all agents accept the contract. Therefore, we integrate Equations~\ref{eq:BR}, \ref{eq:IR1}, and \ref{eq:IR2} as constraints to formulate the principal-MARL contract design problem as follows: 
\begin{center}
\begin{equation}
\begin{aligned}
          \max_{\bm{\alpha}, N_a} \;  \mathcal{G}(\bm{\alpha}, N_a) = \mathbb{E}_{\bm{\pi}^{\bm{\alpha},N_a}}[R^{\bm{\alpha},N_a}]                                                           \nonumber        \\
          s.t. \; \mathbb{E}_{\pi_i^{\bm{\alpha},N_a},\pi_{-i}^{\bm{\alpha},N_a}}[r_i^{\alpha_i}] \geq \mathbb{E}_{{\pi'}_i^{\bm{\alpha},N_a},\pi_{-i}^{\bm{\alpha},N_a}}[r_i^{\alpha_i}],                                       \nonumber\\
         \mathbb{E}_{\pi_j^{\bm{\alpha},N_a},\pi_{-j}^{\bm{\alpha},N_a}}[r_j^{\alpha_j}] - \mathbb{E}_{\pi_j,\pi_{-j}}[r_j] \geq 0   
            \nonumber\\
          \mathbb{E}_{\pi_k^{\bm{\alpha},N_a},\pi_{-k}^{\bm{\alpha},N_a}}[r_k^{\alpha_k}] - c  \geq 0  
            \nonumber\\
          \quad \forall i \in \{1,\cdots,N\}, \forall j \in \{1,\cdots,N_b\}, \forall k \in \{1,\cdots,N_a\}, \\ \forall s \in S, \forall \pi_i, \pi'_{i} \in \Pi_i, \pi_{-i} \in \Pi_{-i}, \nonumber
\end{aligned}
\end{equation}
\end{center}

\noindent where $\mathcal{G}(\bm{\alpha}, N_a)$ is the principal objective and $R^{\bm{\alpha},N_a}$ represents the return from the system given a contract $[\bm{\alpha},N_a]$. The first set of constraints are $|N|$ BRS constraints, ensuring that agents select strategies based on the best response other agents strategies. The second set of constraints are $|N_b|$ IR constraints for the existing agents and the last set of constraints are $|N_a|$ IR constraints for the new agents. We assume new agents have the same the minimum expected return. Note the number of constraints depends on the decision variable $N_a$. 

The expected performance of the system $\mathcal{G}$ under the contract $[\bm{\alpha},N_a]$ is a consequence of the learned joint policy $\pi_i^{\bm{\alpha},N_a},\pi_{-i}^{\bm{\alpha},N_a} \forall i \in N$, obtained by the MARL algorithm. Due to the diverse nature of behaviours that can arise in different contracts and the extensive computational resources needed for multiple training episodes, optimization problems involving MARL are modeled as expensive-to-evaluate black-box functions~\cite{shou2020reward, mguni2019coordinating}. As consequence, solving the principal-MARL contract design problem requires frameworks that can deal with  dynamic number of constraints in the context of expensive-to-evaluate black-box functions.

\section{Methodology}

\begin{figure}[!t]
\centering
\includegraphics[width=3.0in]{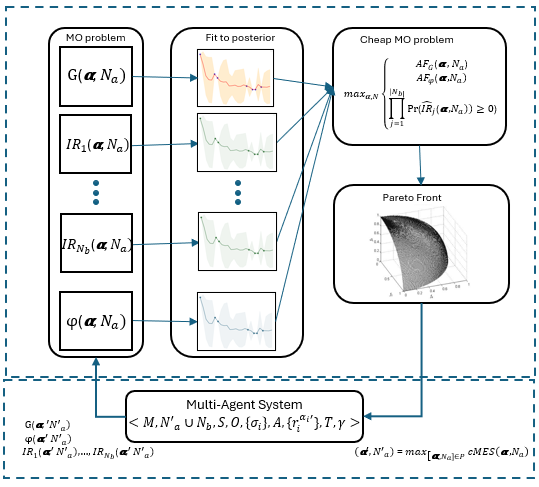}
\caption{cPMES architecture using MOBO to obtain candidates designs at the contract optimization level and the joint policy $\bm{\pi}$ for the given $[\bm{\alpha},N_a]$ at the MARL level.}
\label{fig:cppesframework}
\end{figure}

We propose a novel framework tailored to address the challenging characteristic of the principal-MARL contract design problem, named constrained Pareto Maximum Entropy Search (cPMES). Our framework leverages on independent surrogates for MOBO \cite{Belakaria2020usemo} to overcome the limitations of constrained BO methods and balancing the information gain from the constrains and the probability of feasibility of each contract design. The general framework of cPMES is shown in Figure \ref{fig:cppesframework}, at the contract optimization level we find promising designs $[\bm{\alpha}, N_a]$ and then, at the MARL level, we obtain the system performance $\mathcal{G}(\bm{\alpha}, N_a)$ by computing agents' joint policy for the given contract design.

At the contract optimization level, we set a GP prior over the principal's objective, $|N_b|$ GP priors over the IR constraints of the baselines agents. To address the dynamic number of constraints coming from the IR principle of the additional agents $N_a$, we define a feasibility indicator function as:
\begin{align}
    \phi(\bm{\alpha}, N_a) = 
    \begin{cases} 
        1 & \text{if } \mathbb{E}_{\pi_k^{\bm{\alpha},N_a},\pi_{-k}^{\bm{\alpha},N_a}}[r_k^{\alpha_k}] - c \geq  0, \forall k \in N_a. \\
        0 & \text{otherwise .}\\
    \end{cases} 
\end{align}

The output of the function $\phi(\bm{\alpha}, N_a)$ is independent of the number of agents, therefore, we set a GP prior over the  $\phi$ to account for the feasibility of the IR constraints. By using equilibrium based MARL algorithms, such as Mean Field Actor-Critic~\cite{yangetal2018}, the policies learned by the agents follow the principle of best response strategies, then, it is safe to assume that all policies from the MARL algorithm are feasible from the perspective of the BRS constraints~\cite{mguni2019coordinating}. The proposed modelling leads to a total of $N_b+2$ surrogates for prediction.      

We build a MO problem, using the AF score of the principal's objective $\text{AF}_{\mathcal{G}}(\bm{\alpha}, N_a)$, the AF score of the feasibility indicator function for IR of the additional agents $\text{AF}_{\phi}(\bm{\alpha}, N_a)$ and the product of the predicted feasibility of each IR constraint for the baseline agents, leading to the following multi-objective problem:
\begin{align}
    \max_{\bm{\alpha}, N_a}  \text{AF}_{\mathcal{G}}(\bm{\alpha}, N_a), \text{AF}_{\phi}(\bm{\alpha}, N_a), \prod_{j=1}^{|N_b|} Pr(\hat{IR}_j(\bm{\alpha}, N_a) \geq 0),
    \label{eq:cheapmo}
\end{align}

\noindent where $\hat{IR}_j(\bm{\alpha}, N_a)$ is the surrogate of the IR constraint of the baseline agent $j$, and $Pr(\hat{IR}_j(\bm{\alpha}, N_a) \geq 0)$ is the predicted probability of feasibility of the IR constraint. 

We obtain the Pareto front for the Multi-objective problem~\ref{eq:cheapmo} by using the Non-dominated Sorting Genetic Algorithm II (NSGA-II) \cite{deb2002nsgaii}, a widely used multi-objective evolutionary algorithm that has been applied successfully in the context of MOBO \cite{Belakaria2020usemo}. The Pareto front contains a set of non-dominated solutions, representing the trade-offs between the objectives and showcasing  solutions that offer the best performance across all objectives simultaneously. Therefore, designs that yield a high predicted principal objective but low probability of feasibility are still considered as long as they are non-dominated solutions. To select  designs from the Pareto front $P$, we compute the cMES score, prioritizing the evaluation of designs to maximize the information gain of the principal objective and the IR constraints:
\begin{align}
[\bm{\alpha}', N_a'] = \max_{[\bm{\alpha}, N_a] \in P} \text{cMES}(\bm{\alpha}, N_a).
\label{eq:paretocriteria}
\end{align}

The extension of MOBO frameworks to parallel BO is relatively straightforward~\cite{zhang2022efficient}, cPMES can sample a batch of designs from the Pareto front by following the entropy prediction criteria.  The design or batch of designs selected are taken to the evaluation step at the MARL level. cPMES builds a POMTMG given by the selected design variables $[\bm{\alpha}, N_a]$ as follows:
\begin{align}
    <M, N_a \cup N_b, S, O, \{\sigma_i\}, A, \{r_i^{\alpha_i}\}, T, \gamma>
\end{align}

To solve the POMTMG, we propose Multi Type Mean Field Q-learning (MTMFQ) algorithm~\cite{ganapathi2020multi}. MTMFQ addresses the complexity of MARL by extending the Mean Field Approximation presented in~\cite{yang2018mean} to multiple types of heterogeneous groups of agents. 

The algorithm considers different groups of agents in the environment as types and each agent models its relation to each type separately, selecting the BRS against the mean field action of the different types ~\cite{ganapathi2020multi}. Assuming that the only source of heterogeneity in the POMTMG is given by $\bm{\alpha}$, agents with the same value of $\alpha_i$ can be considered in the same type. 

Once the agents are trained and the policies are evaluated, we update the set of priors for the selected design and repeat the process until the evaluation budget is consumed. The pseudocode for cPMES is shown in Algorithm~\ref{alg:cPPES} of the Appendix~\ref{appendix:theorem}. Our framework provides a systematic and efficient methodology for solving the principal-MARL contract design problem, relying on the feasibility probabilities and entropy-based selection within the Pareto front to enhance the reliability of the proposed contract designs.

A highly desirable property in BO algorithms is to demonstrate that the approach becomes increasingly efficient and effective over time in terms of minimizing cumulative regret \cite{Srinivas2012regbound}. We provide an  upper bound for total regret and discuss the theoretical properties of cPMES. 

Our goal is to establish a sub-linear upper bound for cumulative regret with respect the number of iterations when solving MO problem defined in~\ref{eq:cheapmo}. For the purpose of mathematical derivations, we assume the AF of $\mathcal{G}(\bm{\alpha}, N_a)$ and $\phi(\bm{\alpha}, N_a)$ to be the UCB.  

\textbf{Theorem 1.} Let $[\bm{\alpha}, N_a]^*$ be a solution in the Pareto set $P^*$ and $[\bm{\alpha}, N_a]_t$ a solution in the Pareto set $P_t$ of the MO problem obtained at the $t^{th}$ iteration of cPMES. Let the total regret defined as: 
\begin{align}
    R([\bm{\alpha}, N_a]^*) = \|R_1([\bm{\alpha}, N_a]^*), R_2([\bm{\alpha}, N_a]^*) , R_3([\bm{\alpha}, N_a]^*)\|
\end{align}
\noindent where $\|.\|$ is the norm of the vectors:
\begin{align} 
R_1([\bm{\alpha}, N_a]^*)= \sum^{T_{max}}_{t=1} (\mathcal{G}([\bm{\alpha}, N_a]^*)-\mathcal{G}([\bm{\alpha}, N_a]_t)) \\
R_2([\bm{\alpha}, N_a]^*)= \sum^{T_{max}}_{t=1} (\phi([\bm{\alpha}, N_a]^*)-\phi([\bm{\alpha}, N_a]_t))\\ 
R_3([\bm{\alpha}, N_a]^*)= \sum^{T_{max}}_{t=1} \sum_{j=1}^{|N_b|} [Pr(IR_j([\bm{\alpha}, N_a]^*) \geq 0) \nonumber \\  - Pr(\hat{IR}_j([\bm{\alpha}, N_a]^*) \geq 0)] 
\end{align}
\noindent then, the following holds with probability $1-\gamma$: 
\begin{align}
R([\bm{\alpha}, N_a]^*) \leq \nonumber \\ 
\sqrt{k T_{max}\beta_{T_{max}}(\gamma^{\mathcal{G}}_{T_{max}}+\gamma^{\phi}_{T_{max}} + \sum_{j=1}^{|N_b|}\gamma^{IR_j}_{T_{max}})} 
\end{align}

\noindent where $k$ is a constant and beta is the exploration factor $\beta=2\log((|\bm{\alpha}|+|N_a|)\pi^2t^2/6\delta) $, with $\delta \in [0,1]$ as a confidence parameter. The terms $\gamma^{\mathcal{G}}_{T_{max}}$ and $\gamma^{\phi}_{T_{max}}$ are the maximum information gain about $\mathcal{G}$ and $\phi$ after $T_{max}$ iterations. The proof can be found in Appendix~\ref{appendix:theorem}.

In complex search spaces, such as the principal-MARL contract design, large number of evaluations may be involved. Theorem 1 states that as the number of iterations increases, the regret incurred converges to zero at a sub-linear rate. The sub-linear upper bound underscores the robustness and effectiveness of the approach in complex scenarios, making it a fundamental metric for assessing the practical utility and scalability of the algorithm \cite{Shahriari2016reviewbo}.

\section{Empirical Results}

In this section, we illustrate the application of our proposed method for welfare maximization in the Clean-up problem. We address two distinct case studies: a synthetic version of the Clean-up problem, and the Sequential Social Dilemma (SSD) version with dynamic environment. As our work showcases the introduction of the principal-MARL contract design problem, we conduct benchmark experiments by replacing the MOBO portion of our approach with relevant constrained BO methods, cEI, cMES and MACE. Furthermore, we perform parallel BO experiments to test the batch performance of cPMES against MACE with batch sizes 2 and 4. 

The Clean-up environment, introduced by \cite{hughes2018inequity}, is a SSD problem for MARL, which focuses on achieving welfare maximization in multi-agent systems with social dilemma properties \cite{leibo2017multi}.  The environment is split in two areas, river area and the harvesting area. Agents  can harvest apples from the harvesting area, apple will continue to re spawn as long as the river is clean. The river becomes progressively polluted, so it requires for agents to clean it to ensure the long term availability of apples. Harvesting apples yields a positive reward, while cleaning the river does not yield a reward from the environment. 

We define an instance of the clean-up problem where agents are restricted to two categories, harvesters and cleaners. We assume principal has a system of $N_b$ harvesters and he/she would like to improve the system welfare by recruiting $N_a$ cleaners. Since cleaning the river does not yield an intrinsic reward, principal wants to introduce a tax $\alpha$ to the harvesters' income and redistribute equally among cleaners. The principal-MARL contract design for the clean-up problem is defined by:
\begin{align}
         \max_{\alpha, N_a} \;  \mathcal{G}(\alpha, N_a) = \mathbb{E}_{\bm{\pi}^{\alpha,N_a}}[\sum_{j=1}^{|N_b|}r_j-c_j - \sum_{k=1}^{|N_a|}c_k ]                                                           \nonumber        \\
         s.t. \; \mathbb{E}_{\pi_j^{\alpha,N_a},\pi_{-j}^{\alpha,N_a}}[(1-\alpha)*r_j- c_j] \geq \nonumber \\ 
         \mathbb{E}_{{\pi'}_j^{\alpha,N_a},\pi_{-j}^{\alpha,N_a}}[(1-\alpha)*r_j- c_j],                                       \nonumber\\
         \mathbb{E}_{\pi_k^{\alpha,N_a},\pi_{-k}^{\alpha,N_a}}[\frac{\alpha}{|N_a|} \sum_{j=1}^{|N_b|}r_j - c_k] \geq \nonumber \\ 
         \mathbb{E}_{{\pi'}_k^{\alpha,N_a},\pi_{-k}^{\alpha,N_a}}[\frac{\alpha}{|N_a|} \sum_{j=1}^{|N_b|}r_j - c_k],                                       \nonumber\\
       \mathbb{E}_{\pi_j^{\alpha,N_a},\pi_{-j}^{\alpha,N_a}}[(1-\alpha)*r_j - c_j] - \mathbb{E}_{\pi_j,\pi_{-j}}[r_j- c_j] \geq 0   
            \nonumber\\
       \mathbb{E}_{\pi_k^{\alpha,N_a},\pi_{-k}^{\alpha,N_a}}[\frac{\alpha}{|N_a|} \sum_{j=1}^{|N_b|}r_j - c_k] - c  \geq 0  
            \nonumber\\
        \quad \forall i \in \{1,\cdots,N\}, \forall j \in \{1,\cdots,N_b\}, \forall k \in \{1,\cdots,N_a\},\nonumber \\
         \forall s \in S, \forall \pi_i, \pi'_{i} \in \Pi_i, \pi_{-i} \in \Pi_{-i}, \nonumber
\end{align}

\noindent where $c_j$ and $c_k$ are the cost function for harvesters and cleaners, respectively. The principal objective $\mathcal{G}(\alpha, N_a)$ is to maximize the welfare of the system, express as the summation of intrinsic reward for harvesting minus agents' costs. Note that $\alpha$ indirectly affects the principal objective trough the learned policy of the agents in interaction with the environment. Such complex setting can not be obtained in traditional principal-multi-agent problems. 

We fit the principal objective $\mathcal{G}$ and the IR constraints for harvesters $\hat{IR}_j \forall j \in N_b$ to a GP with mean 0 and a kernel function composed by the product of 2 squared exponential (SE) kernels, which works well in discrete search with multiple design variables \cite{duvenaud2014kernel}. The feasibility indicator function $\phi$ is fit to a GP with mean 0 and  a kernel function composed of the product of 2 Matérn kernels and a constant kernel to capture the variability of the binary function. 

\subsection{Synthetic Problem}

In this set of experiments, we focus on the performance at the design level, bypassing the MARL level by randomly assigning a performance value to each contract design, creating a known performance distribution. Further implementation details can be found in Appendix~\ref{appendix:algorithm}.  This setting enables us to benchmark the empirical regret of each method, by comparing the best-suggested contract design within a fixed budget of evaluations against the truly optimal contract. 

We measure regret as the absolute difference of the principal objective between the best contract design found in each method and the optimal contract under five different budgets: $T_{max}=4$,$T_{max}=8$,$T_{max}=12$,  $T_{max}=16$, $T_{max}=20$. We consider $|N_b| = 5$ and the principal can recruit up to 3 cleaners. We setup five random seeds, each prior set has the lower and upper bound  of combinations of design variable and three randomly selected designs, as in ~\cite{shou2020reward}.  We report in Table~\ref{tab:resultsSC} the average regret obtained for each methods across the set of five different priors. In Table~\ref{tab:resultsSB} we report the results for parallel experiments in batch sizes of $B=2$ and $B=4$.

\begin{table}[!t]
  \caption{Average regret and 95\% confidence interval for each evaluation budget.}
  \label{tab:resultsSC}
  \centering
  \setlength{\tabcolsep}{1mm} 
  \begin{tabular}{|p{0.9cm}|*{4}{p{1.5cm}|}} 
    \hline
    $T_{\text{max}}$ & \textit{cEI} & \textit{cMES} & \textit{MACE} & \textit{cPMES} \\
    \hline
    $4$ & $1891.4 \pm 1917.5$ & $1448.4 \pm 1497.5$ & $745.0 \pm 512.0$ & $1308.2 \pm 1624.9$ \\
    \hline
    $8$ & $1891.4 \pm 1917.5$ & $686.0 \pm 399.6$ & $578.2 \pm 197.1$ & $1075.6 \pm 1689.6$ \\
    \hline
    $12$ & $1891.4 \pm 1917.5$ & $449.8 \pm 372.9$ & $578.2 \pm 197.1$ & $465.4 \pm 80.4$ \\
    \hline
    $16$ & $1891.4 \pm 1917.5$ & $449.8 \pm 372.9$ & $578.2 \pm 197.1$ & $465.4 \pm 80.4$ \\
    \hline
    $20$ & $1891.4 \pm 1917.5$ & $449.8 \pm 372.9$ & $578.2 \pm 197.1$ & $375.0 \pm 233.0$ \\
    \hline
  \end{tabular}
\end{table}

\begin{table}[!t]
  \caption{Average regret and 95\% confidence interval of parallel methods for each evaluation budget.}
  \label{tab:resultsSB}
  \centering
  \setlength{\tabcolsep}{0.5mm} 
  \begin{tabular}{|p{0.9cm}|*{4}{p{1.5cm}|}} 
    \hline
    \multirow{2}{*}{$T_{\text{max}}$} & \multicolumn{2}{c|}{\textit{Batch Size} $B=2$} & \multicolumn{2}{c|}{\textit{Batch Size} $B=4$} \\
    \cline{2-5}
    & \textit{MACE} & \textit{cPMES} & \textit{MACE} & \textit{cPMES} \\
    \hline
    $4$ & $880.2 \pm 472.2$ &  $1404.8 \pm 1536.9$ & $940.4 \pm 448.6$ &  $1404.8 \pm 1536.9$ \\
    \hline
    $8$ & $869.2 \pm 443.3$ &  $772.8 \pm 560.1$ &  $902.2 \pm 472.0$ & $574.0 \pm 179.4$ \\
    \hline
    $12$& $791.6 \pm 405.3$ & $772.8 \pm 560.1$ &  $788.0 \pm 371.6$ & $574.0 \pm 179.4$\\
    \hline
    $16$ & $773.0 \pm 405.0$  & $718.8 \pm 618.6$ & $755.2 \pm 376.4$ & $574.0 \pm 179.4$ \\
    \hline
    $20$ & $594.6 \pm 591.6$  & $628.6 \pm 429.4$ & $698.6 \pm 430.3$ & $530.4 \pm 203.2$ \\
    \hline
  \end{tabular}
\end{table}

The highly constrained nature of the principal-MARL contract design problem significantly affects the search for feasible solutions, with most methods struggling to improve the regret in the first 8 iterations. cPMES systematically reduces the regret as the number of evaluations is increasing, achieves the lowest regret across all methods and demonstrating promising results in this domain. Conversely, cEI fails to effectively reduce regret as its acquisition function score overfits the best feasible design in the set of priors in highly constrained problems. cMES reduces regret at a slower pace than cPMES, however it maintains a high variance due to the entropy score promoting designs that safe in terms of feasibility. MACE struggles to find better feasible designs due to heavy penalization of predicted infeasible designs and a random selection rule that does not promote information gain.

Benchmark studies for parallel approaches show that cPMES with a batch size of four achieves the lowest regret across experiments. MACE performs best with $B=2$, reducing the regret to 594.6 in average. Designs with higher entropy score tend to be clustered in the distribution, providing similar information to the GP. In the other hand, MACE's random selection provides a more diverse selection of designs to the batch. The clustering issue in cPMES becomes less relevant as the batch size increases.  

The results illustrate the efficacy of each constrained BO method in navigating the highly constrained design space of the synthetic problem. We observe that cPMES leverages MOBO more effectively than MACE with batch size four, achieving minimum regret in 8 or fewer iterations. Further experiments in MARL can test the robustness of these methods. The promising results obtained with cPMES are further validated in a dynamic environment in the subsequent section.

\subsection{Sequential Social Dilemma}

We present the implementation of the SSD Clean-up for MARL. We consider an environment with five baseline harvesters and we allow principal to recruit up to five cleaners. The variable $\alpha$ acts as a tax for the harvesters income, which is equally distributed among cleaners, regardless of their performance.  The general layout and implementation details can be found in Appendix~\ref{appendix:algorithm}. 

The experimental setup involves analyzing the joint behavior of agents under different contract designs assuming a minimum expected return for cleaners of $0$. The outcome for the principal is highly dependent on the environment dynamics, leading to a richer analysis compared to the static setup of traditional contract designs. By employing RL agents, we can perform dynamic and adaptive analyses that are not possible in static settings. For instance, we can evaluate how agents adapt their strategies over time, respond to changing environments, and react to incentives or penalties. This approach allows us to study systems that more closely resemble real-world dynamics.

\subsubsection{Results}

\begin{figure}[!t]
\centering
\includegraphics[width=3.0in]{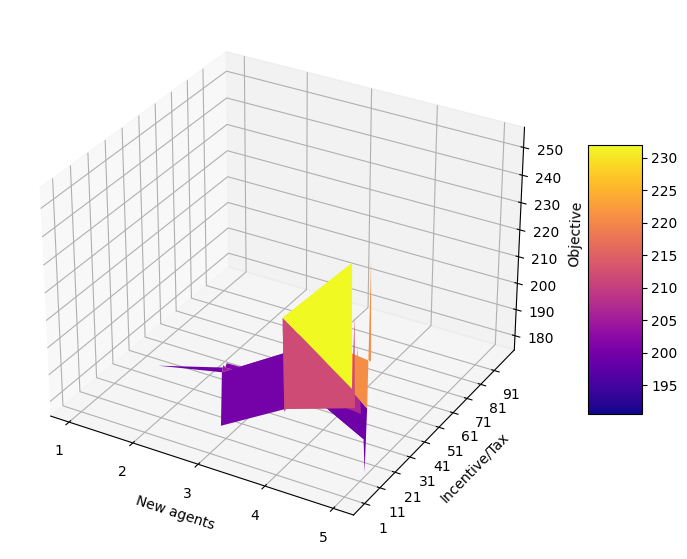}
\caption{Principal's objective and recommended feasible contracts}
\label{fig:resultsmarl1}
\end{figure}

We conduct experiments of cEI, cMES, MACE, and cPMES, across a series of 20 evaluations and using a set of 10 initial priors across five different random seeds. The  Figure~\ref{fig:resultsmarl1} shows a heatmap of all the feasible designs obtained across all experiments. Most methods aim for designs with the highest number of new agents and low $\alpha$ to avoid violating any of the IR constraints. The best contract design found is $[N_a=5,\alpha= 0.04)$, providing a Principal's objective of $255.53$, cleaners utility of $0.28$ and each harvester obtains an expected return over $49.96$, while in the baseline situation, each harvester gets an utility no higher than $30.27$.

\begin{figure}[htbp]
\centering

\begin{minipage}[t]{0.22\textwidth}
    \centering
    \includegraphics[width=\textwidth]{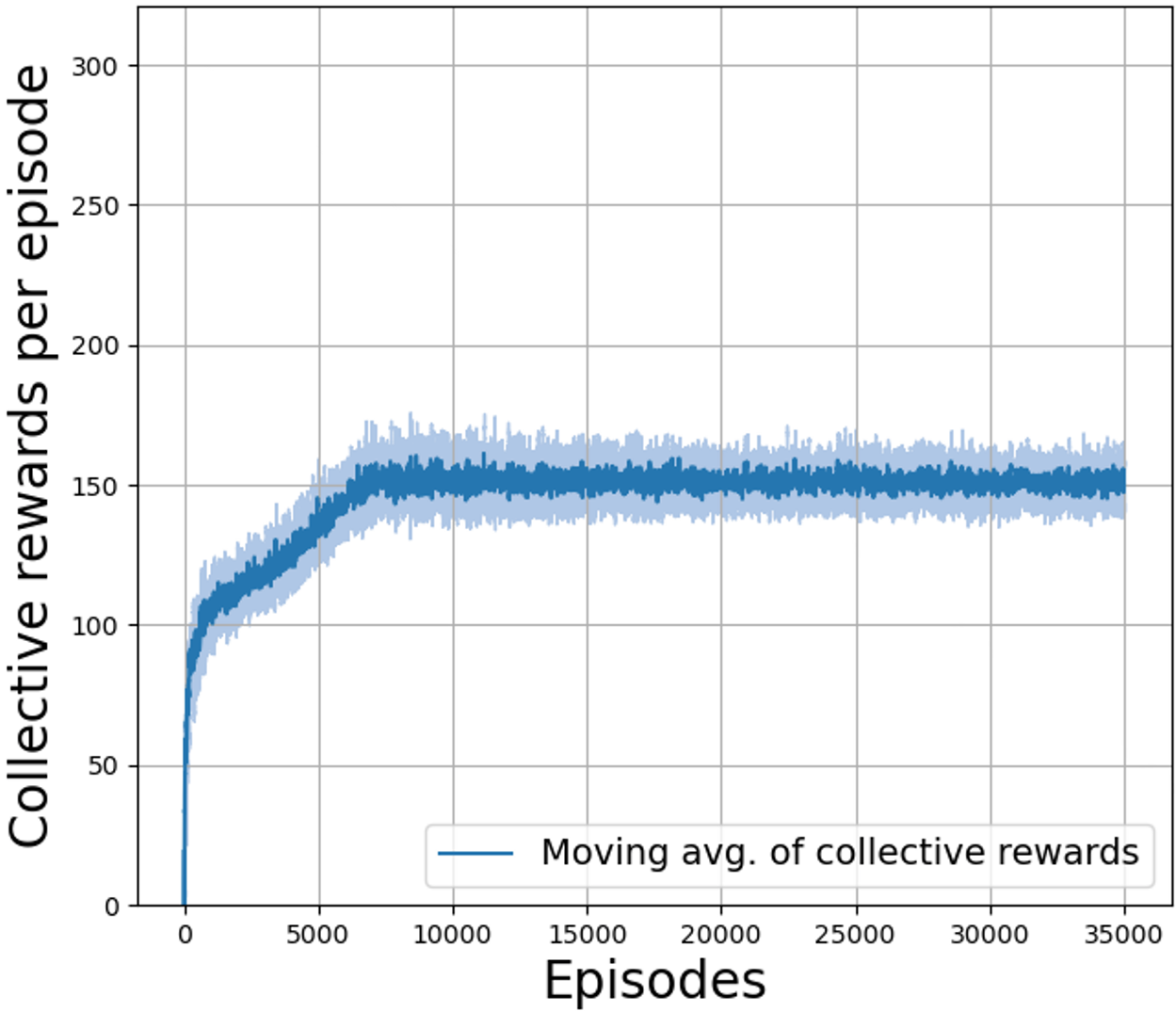}
    \caption{Training performance $N_a=1$ and $\alpha=0.01$}
    \label{fig:sub1}
\end{minipage}%
\hfill
\begin{minipage}[t]{0.22\textwidth}
    \centering
    \includegraphics[width=\textwidth]{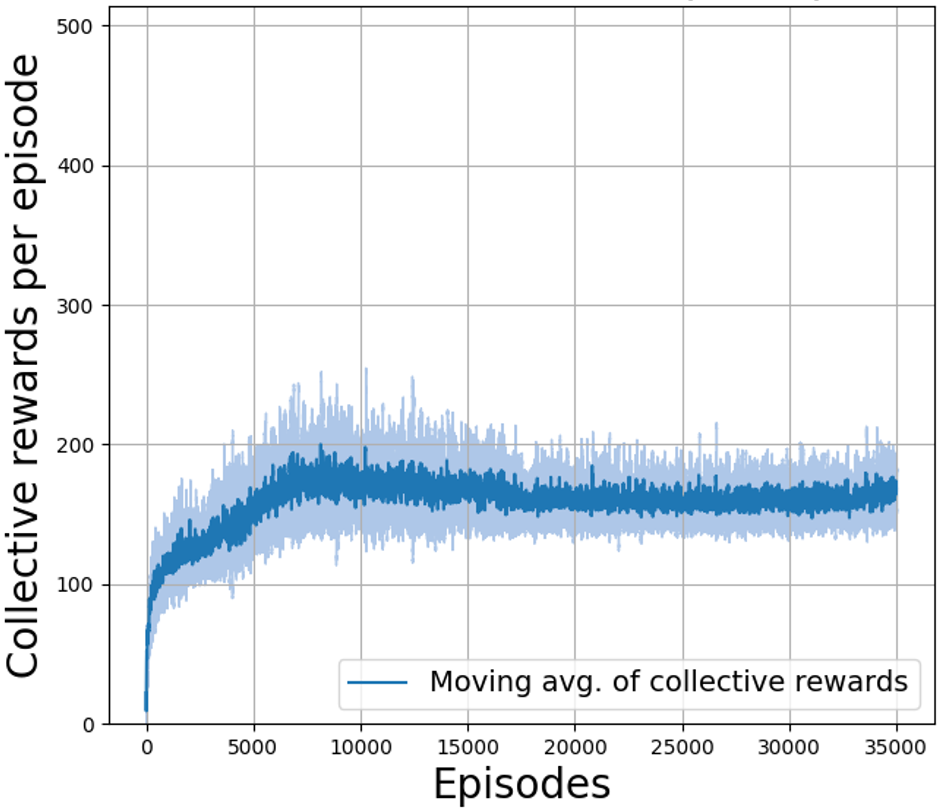}
    \caption{Training performance $N_a=5$ and $\alpha=0.01$}
    \label{fig:sub2}
\end{minipage}

\vspace{1em}

\begin{minipage}[t]{0.22\textwidth}
    \centering
    \includegraphics[width=\textwidth]{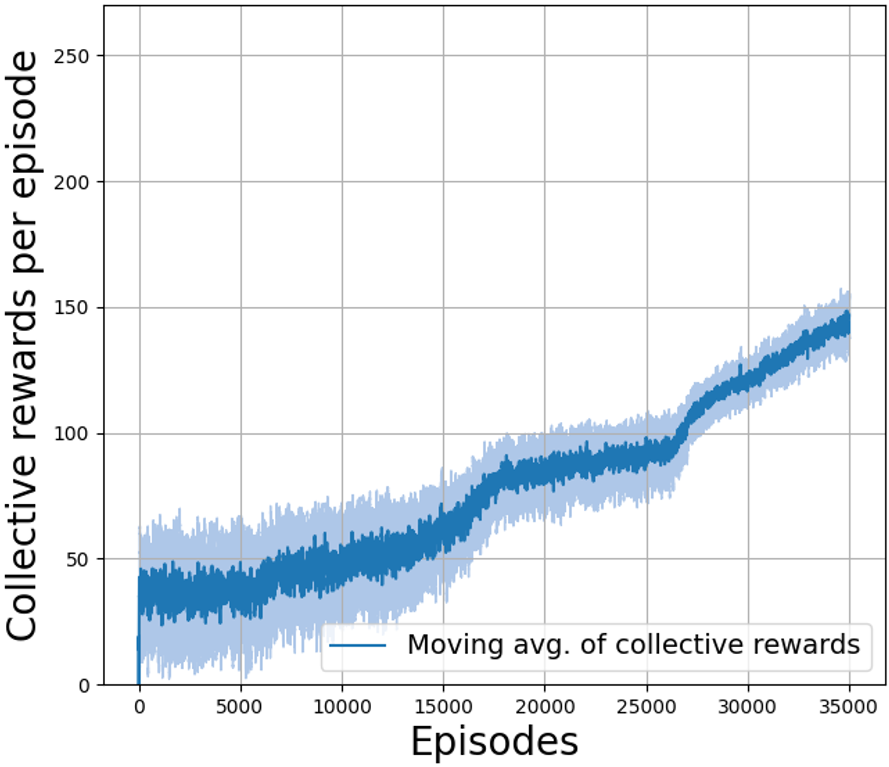}
    \caption{Training performance $N_a=1$ and $\alpha=0.99$}
    \label{fig:sub3}
\end{minipage}%
\hfill
\begin{minipage}[t]{0.22\textwidth}
    \centering
    \includegraphics[width=\textwidth]{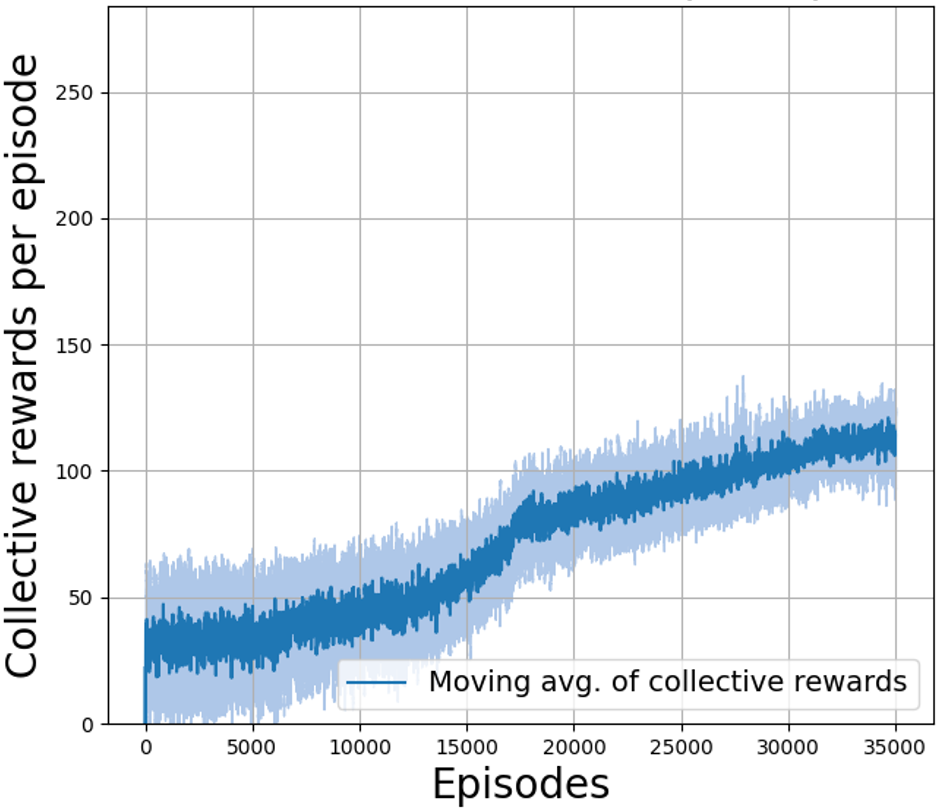}
    \caption{Training performance $N_a=5$ and $\alpha=0.99$}
    \label{fig:sub4}
\end{minipage}

\caption{Training performance under different contract designs.}
\label{fig:global}
\end{figure}

We provide an analysis of how the learning of the multi-agent system is affected by the contract design. In Figure~\ref{fig:global}, we present the training curves of four different contracts, each varying in the number of recruited agents $N_a$ and the harvest tax $\alpha$. The collective rewards per episode are plotted over a series of 35,000 episodes for each contract design.

In the scenario presented in Sub-figure~\ref{fig:sub1}, the system shows a rapid increase in collective rewards during the initial episodes, then the moving average of collective rewards stabilizes around 150. The curve indicates that a adding a single cleaner with a low $\alpha$ quickly converges to a joint strategy with low variance across the training. The Sub-figure~\ref{fig:sub2} showcases five additional agents and the same low tax, the system exhibits a slower convergence and more variance compared to recruiting a single-agent. However, it achieves a moving average of collective rewards around 225. This suggest that the presence of additional agents increase the complexity of learning best response strategies due to multiple equilibria. 

In the case of single cleaner agent with a high $\alpha$ shown in Sub-figure~\ref{fig:sub3}, the training curve shows a more volatile pattern. Despite the fluctuations, there is a clear upward trend, with the moving average of collective rewards steadily increasing throughout the training period, reaching approximately 100 by 35,000 episodes. For five additional agents with a high tax, showcased in Sub-figure~\ref{fig:sub4}, the training performance is characterized by gradual gains with higher variability. This indicates that contracts with high value of $\alpha$ require more episodes to converge to a stable performance, due to the fact that the wealth generator agents, the harvester, receive less feedback from the environment. making it harder for learn effective strategies. 

\begin{figure}[!t]
\centering
\includegraphics[width=3.5in]{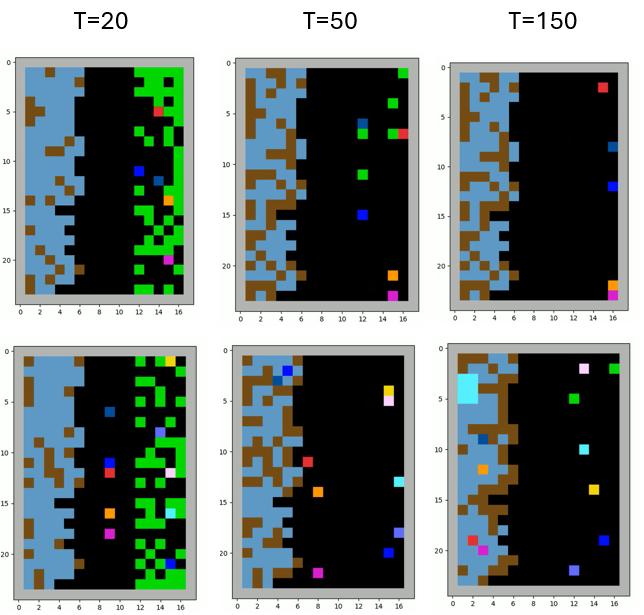}
\caption{Comparison of agents behaviours between baseline setup (upper row) and best contract design (lower row).}
\label{fig:resultsmarlb}
\end{figure}

The comparison of the behaviors obtained in the best recommended contract design with the baseline setting, which only considers harvester agents, is shown in Figure~\ref{fig:resultsmarlb}. Initially, the environment has plenty of apples to be harvested. However, around time-step 50, the apple spawn rate is reduced due to the accumulation of waste in the river. At this point, we observe that the harvesting behavior under the best contract design is more efficient, leading to faster depletion of apples compared to the same number of harvesters in the baseline setup. By time-step 150, the baseline setup shows complete depletion of apples, whereas the best contract design maintains apple spawning thanks to the presence of cleaners in the river. The policy learned by the harvesters under the cPMES contract design is superior, as the agents have experienced more instances of apple collection than the baseline harvesters, due to the additional five cleaners. 

\section{Conclusions}

We present a novel problem setting for algorithmic contract design, the principal-MARL contract design problem. We propose a MOBO-based approach to optimize the contract design and a MARL algorithm to train the agents and evaluate the system performance. Our method, cPMES, integrates the probability of feasibility of the designs as an objective to include promising non-dominated solutions in the border of feasibility and selecting solutions based on the predicted information gain. We show the capabilities of cPMES to navigate highly constraint search space, effectively finding feasible contract designs to maximize principal objective.  

Applications of cPMES are not only limited to the principal-MARL contract design and it can extended to highly constraint with expensive-to-evaluate black-box problem in other domains. A key area for future research lies in addressing the primary bottleneck for solving the principal-MARL contract design problem—training agents using MARL algorithms. Introducing a hyper-network that can generalize policies over different combination of design variables could offer a promising avenue to mitigate this bottleneck, enabling the use of generalized policies across different contract designs and reducing the computational burden of the optimization process.

\clearpage
\end{document}